\documentclass[sigconf]{acmart}

\AtBeginDocument{%
  \providecommand\BibTeX{{%
    \normalfont B\kern-0.5em{\scshape i\kern-0.25em b}\kern-0.8em\TeX}}}

\renewcommand\footnotetextcopyrightpermission[1]{}

\usepackage{adjustbox}
        
\usepackage{amsmath,amssymb,amsfonts}
\usepackage[linesnumbered,ruled,vlined]{algorithm2e}
\usepackage{setspace}
\usepackage{algorithmic}
\usepackage{float}
\usepackage{graphicx}
\usepackage{textcomp}
\usepackage{xcolor}
\usepackage{multirow}
\usepackage{mdframed}
\usepackage{balance}
\usepackage{booktabs}
\usepackage{tcolorbox}
\usepackage{makecell}
\usepackage{threeparttable}
\usepackage{colortbl}
\usepackage{subfigure}
\usepackage{hhline}
\usepackage{pifont}
\usepackage{listings}
\usepackage{enumitem}

\newcommand{\parabf}[1]{\noindent\textbf{#1}}

\definecolor{ggray}{HTML}{eff0f0}
\definecolor{gggray}{HTML}{E8E8E8}
\definecolor{ggggray}{HTML}{BEBEBE}
\definecolor{myblue}{RGB}{255,255,255}
\definecolor{myyellow}{HTML}{FFF2CC}

\newcommand{\ie}{\textit{i.e.,}}
\newcommand{\eg}{\textit{e.g.,}}

\newcommand{\ourtool}{STALL$^+$}
\newcommand{\repocoder}{RepoCoder}

\newcommand{\mgd}{MGD}
\newcommand{\rlpg}{RLPG}
\newcommand{\cocomic}{CoCoMIC}

\newcounter{finding}
\newcommand{\finding}[1]{\refstepcounter{finding}
 	\vspace{1mm}
	\begin{mdframed}[linecolor=gray,roundcorner=12pt,backgroundcolor=gray!15,linewidth=3pt,innerleftmargin=2pt, leftmargin=0cm,rightmargin=0cm,topline=false,bottomline=false,rightline = false]
		\textbf{Answer to RQ\arabic{finding}:} #1
	\end{mdframed}
	\vspace{3mm}
}

\newcommand{\distance}{5pt}
\newcommand{\rg}{RAG}
\newcommand{\infile}{In-file}
\newcommand{\si}{Prompt-F}
\newcommand{\sls}{Prompt-T}
\newcommand{\spo}{Post}
\newcommand{\sd}{Decode}
\newcommand{\sid}{\si{} + \sd{}}
\newcommand{\sip}{\si{} + \spo{}}
\newcommand{\slsd}{\sls{} + \sd{}}
\newcommand{\slsp}{\sls{} + \spo{}}
\newcommand{\rsi}{\rg{} + \si{}}
\newcommand{\rsls}{\rg{} + \sls{}}
\newcommand{\rsp}{\rg{} + \spo{}}
\newcommand{\rsd}{\rg{} + \sd{}}
\newcommand{\rsid}{\rg{} + \sid{}}
\newcommand{\rsip}{\rg{} + \sip{}}
\newcommand{\rslsd}{\rg{} + \slsd{}}
\newcommand{\rslsp}{\rg{} + \slsp{}}
\newcommand{\starcoder}{StarCoderBase-7B}
\newcommand{\codellama}{CodeLlama-7B}
\newcommand{\deepseekcoder}{DeepSeek-Coder-6.7B}
\newcommand{\lineem}{line EM}

\newcommand{\importcontext}{File-level Dependency}
\newcommand{\apicontext}{Token-level Dependency}

\newcommand{\cceval}{CrossCodeEval}

\definecolor{myorgange}{RGB}{251,229,214}
\definecolor{mygreen}{RGB}{226,240,217}
\definecolor{myblue}{RGB}{218,227,243}
\definecolor{mygray}{RGB}{230,230,230}

\usepackage{soul}

\usepackage{multirow}
\usepackage[normalem]{ulem}
\useunder{\uline}{\ul}{}

\newcommand{\bestres}[1]{{\ul \textbf{#1}}}

\begin{document}

\title{STALL$^+$: Boosting LLM-based Repository-level Code Completion with Static Analysis}

\author{Junwei Liu}
\affiliation{%
  \institution{Fudan University}
  \city{Shanghai}
  \country{China}
  }

\author{Yixuan Chen}
\affiliation{%
  \institution{Fudan University}
  \city{Shanghai}
  \country{China}
  } 

\author{Mingwei Liu}
\affiliation{%
  \institution{Sun Yat-sen University}
  \city{Guangzhou}
  \country{China}
  }
  
\author{Xin Peng}
\affiliation{%
  \institution{Fudan University}
  \city{Shanghai}
  \country{China}
  }  
  
\author{Yiling Lou}
\affiliation{%
  \institution{Fudan University}
  \city{Shanghai}
  \country{China}
  }

\begin{abstract}
Repository-level code completion is challenging as it involves complicated contexts from multiple files in the repository. To date, researchers have proposed two technical categories to enhance LLM-based repository-level code completion, \ie{} retrieval-augmented generation (RAG) and static analysis integration. 
This work performs the first study on the static analysis integration in LLM-based repository-level code completion by investigating both the effectiveness and efficiency of static analysis integration strategies across different phases of code completion. We first implement a framework~\ourtool{}, which supports an extendable and customizable integration of multiple static analysis strategies into the complete pipeline of LLM-based repository-level code completion; and 
based on  \ourtool{}, we perform extensive experiments by including different code LLMs on the latest repository-level code completion benchmark \cceval{}. Our findings show that integrating file-level dependencies in  prompting phase performs the best while the integration in post-processing phase performs the worse. Additionally, we  observe different improvements from static analysis between dynamic languages and static languages, \ie{} the best combination is prompting-phase  with decoding-phase integration for Java while the best combination is prompting-phase with post-processing-phase integration for Python given the limitations of statically analyzing dynamic languages. Additionally, we find the complementarity between RAG and static analysis integration as well as their cost-effectiveness after combination.
\end{abstract}
\maketitle

\section{Introduction}
Code completion techniques automatically generate code for given contexts (\eg{} natural language descriptions or incomplete code snippets), which have been widely adopted in practical programming activities and substantially increase development productivity~\cite{li2021toward,wang2023practitioners,DBLP:journals/corr/abs-2401-00288}. Recent advances in deep learning, especially large language models (LLMs), further boost the progress in this domain~\cite{DBLP:journals/corr/abs-2312-15223, DBLP:journals/corr/abs-2308-10620, DBLP:conf/fose-ws/FanGHLSYZ23}. 
Having been pre-trained on massive code corpus, code LLMs exhibit strong effectiveness in code completion. 

Previously, the majority of code completion work~\cite{classeval,ni2023lever,pmlr-v202-guo23j} focuses on code completion within a small-scale and isolated context (\ie{} a single file). For example, coding tasks in widely-used benchmarks (such as HumanEval~\cite{HumanEval} and MBPP~\cite{MBPP}) expect to generate code mainly based on the context of the current file. 
However, real-world code completion is often associated with a broader repository-level context. In such scenarios (\textbf{repository-level code completion}), the code in the target file (\ie{} the file with unfinished code for completion) may depend on code contexts (\eg{} classes or methods) declared in other files of the repository. 
Unlike previous code completion only taking the target file as input, repository-level code completion takes not only the target file but also the entire repository as input. 

Compared to code completion within a single file, using LLMs for repository-level code completion is more challenging. First, generating code dependent on other files of the repository is a more restricted task for LLMs. 
While most LLMs are trained on public code corpus like GitHub projects, they may lack knowledge of the token distribution within the local repository. Therefore, LLMs are more likely to exhibit hallucination when generating code tokens that are less commonly seen in their pre-training data. Previous work~\cite{cceval}  shows that state-of-the-art code LLMs perform much poorer when generating code dependent on other local files.  Second, the overwhelming scale of the repository makes it infeasible to feed the complete repository-level context into LLMs due to their limited input length and poorer performance with longer inputs~\cite{liu2024lost}. As a result, only a portion of repository-level context can be utilized by the LLMs, which potentially leads to the non-optimal effectiveness. 

To address the issues above, researchers have proposed two categories of techniques to boost LLM-based repository-level code completion. The first category is to enhance LLM-based repository-level code completion via Retrieval-Augmented Generation (RAG) techniques~\cite{repocoder, repoformer}. In particular, RAG-based techniques first leverage a retrieval mechanism to fetch similar code snippets from the repository, and then include the retrieved code snippets in the input prompt. The prompts augmented with the similar code examples facilitate more accurate LLM-based code completion given the idea of few-shot learning and in-context learning~\cite{gpt3}. The second category is to enhance LLM-based repository-level code completion with static analysis. In particular, existing techniques mainly integrate static analysis in two phases, \ie{} (i) integrating static analysis in the prompting phase by extracting useful code contexts from other files via static analysis~\cite{cocomic, RepoBench} and (ii) integrating static analysis in the decoding phase by adjusting the model-predicted token probabilities guided by static analysis ~\cite{mgd}. While the RAG category enhances LLMs with similar code in the repository, static analysis equips LLMs with additional capabilities of code analysis.  

In this work, we investigate the static analysis integration in LLM-based repository-level code completion by investigating both the effectiveness and efficiency of static analysis integration strategies across different phases of code completion. In particular, we make the first attempt to (i) compare the effectiveness and efficiency of each individual integration strategy, (ii) the complementarity among different strategies based on their combination, and (iii) the comparison and combination to existing RAG techniques. 

To facilitate such a comprehensive evaluation, we first propose a framework~\ourtool{}, which supports an extendable and customizable integration of multiple static analysis strategies into the complete pipeline of LLM-based repository-level code completion. In particular, \ourtool{} can integrate static analysis along the prompting phase (before model inference), the decoding phase (during model inference), and the post-processing phase (after model inference). Additionally, \ourtool{} is not only extendable for different static analysis strategies and their combination, but also compatible with RAG techniques. 

Based on our framework \ourtool{}, we perform extensive experiments by including three state-of-the-art code LLMs (\ie{} \deepseekcoder{}, \starcoder{}, and \codellama{}) on the latest repository-level code completion benchmark \cceval{}, which includes 2,075/2,460 code completion tasks of both static language Java and dynamic language Python. Based on our experimental results, we have the following main findings. First, integrating static analysis in any phase of code completion can improve repository-level code completion, while prompting phase with file-level dependencies performs the best and the post-processing phase performs the worse. Second, combining multiple integration strategies can bring further improvements, and different strategies exhibit different complementarity (\eg{} the prompting phase with token-level dependencies shares the smallest complementarity with decoding-phase integration). In particular, integrating static analysis for dynamic language or static language exhibit different improvements, \eg{} the best combination is prompting-phase with decoding-phase for Java, while the best combination is prompting-phase with post-processing-phase for Python given the limitations of statically analyzing dynamic languages. Third, static analysis integration outperforms RAG in repository-level code completion, and combining them can further achieves the best accuracy. Forth, integrating static analysis in the prompting phase is the most efficient way, while combining prompting-phase static analysis and RAG is the best option for cost-effectiveness. Additionally, we further discuss practical implications on static analysis integration strategies for the future work, \eg{} more flexible and  efficient integration strategies.

In summary, this work makes the following contributions:

\begin{itemize}[leftmargin=15pt, topsep=2pt]

\item We investigate both the effectiveness and efficiency of static analysis integration strategies across different phases of LLM-based repository-level code completion.

\item We propose and implement a framework~\ourtool{}, 
 which supports an extendable and customizable integration of multiple static analysis strategies into the complete pipeline of LLM-based repository-level code completion.

\item Our study reveals many findings and practical implications on static analysis integration, including the best strategy, the most cost-effectiveness strategy, the combination and complementary between static analysis and RAG, and the potential future directions for designing more powerful static analysis integration strategies. 
\end{itemize}

\section{Background}

\subsection{Code LLMs}
Large language models (LLMs) such as Llama~\cite{llama} and ChatGPT~\cite{chatgpt}, are large-scale Transformer models with hundreds of billions of parameters. Recently LLMs have made significant progress and been applied across various domains. Code LLMs are specifically trained on code corpus, such as GitHub repositories, showcasing strong capabilities in code generation and code comprehension. Currently, many LLMs have been released, including CodeGen~\cite{codegen}, StarCoder~\cite{starcoder}, and Code Llama~\cite{codellama}, which have been widely applied in code-related tasks, such as code completion~\cite{eval_instruct,classeval,dou2024stepcoder}, test generation~\cite{codamosa, Deng2023LLMFuzzers}, and program repair~\cite{Bouzenia2024RepairAgentAA, Chow2024PyTyRS, SkipAnalyzer}.

\subsection{Repository-level Code Completion}
Repository-level code completion~\cite{cceval, repocoder, mgd} focuses on completing unfinished code based on the context of the entire repository. Unlike traditional code completion tasks that consider only the target file (\ie{} the file with unfinished code to be completed), repository-level code completion takes the entire repository as input. The output typically consists of completed code that may depend on other files within the repository, such as invoking methods declared in other files.
Compared to code completion within a small-scale and isolated context (\eg{} within a single file~\cite{HumanEval, MBPP, classeval}), repository-level code completion is considered a more realistic scenario, as files in real-world software development often depend on each other.

However, repository-level code completion has shown to be more challenging than code completion within the single-file context~\cite{cceval}. Researchers have proposed various techniques to enhance LLMs performance in repository-level code completion, which broadly fall into two categories: (i) Retrieval-Augmented Generation (RAG)~\cite{repocoder, repoformer} and (ii) static analysis integration~\cite{mgd}.

\textit{RAG-based techniques} focus on the prompting phase of LLM-based code completion. They retrieve code snippets that are similar to the unfinished one from the repository and include them in the input prompt. The main insight of RAG-based techniques is to leverage the few-shot learning and in-context learning capabilities of models by providing similar code examples as hints~\cite{gpt3}. However, when no similar code snippets are available in the repository (\ie{} the code to be complete is almost unique), RAG-based techniques can become less effective~\cite{cceval}. 

\textit{Static analysis integration techniques} enhance LLMs with the code analyzing and checking capabilities of static analysis. Existing techniques mainly integrate static analysis in two phases: (i) the prompting phase (before the model inference), which utilizes static analysis to extract useful/dependent code contexts from other files and then prepends them into the current context~\cite{cocomic, RepoBench}; (ii) the decoding phase (during the model inference), which utilizes static analysis to extract valid tokens (\eg{} method names) on the fly and then based on them adjusts the model-predicted token probabilities~\cite{mgd}. This category enhances LLMs by equipping them with additional capabilities from external tools (\ie{} static analyzers).

\begin{figure*}[htb]
	\centering
    \vspace{-3mm}	\includegraphics[width=2.0\columnwidth]{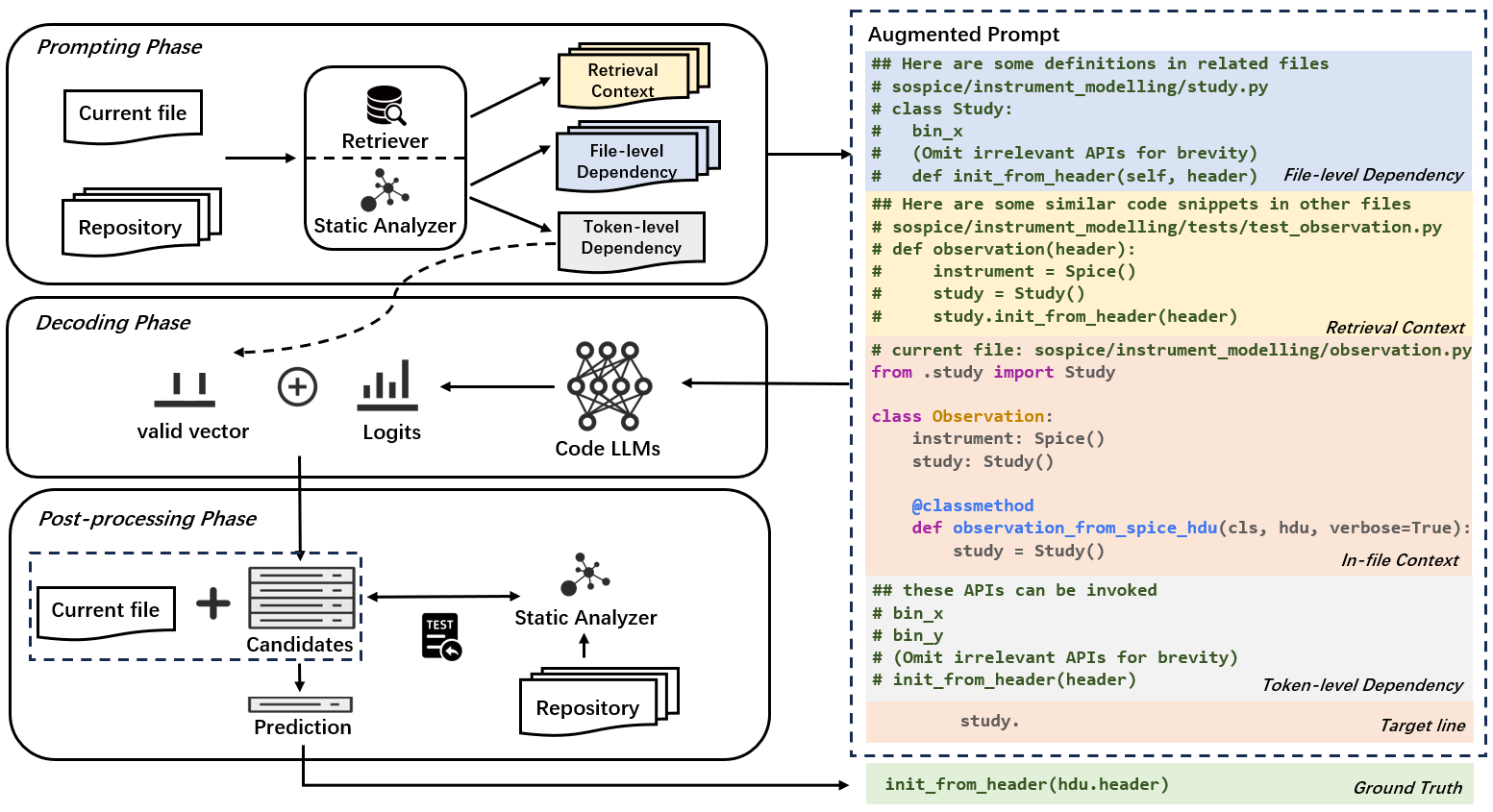}
	\caption{Overview of the \ourtool{} static analysis integration framework
 } 
	\label{fig:overview}
 \vspace{-2mm}
\end{figure*}
\section{Integration Framework for Static Analysis}\label{sec:framework}
In this section, we present our framework, \ourtool{}, which enables an extendable and customizable integration of static analysis throughout the pipeline of LLM-based repository-level code completion. 
As shown in Figure~\ref{fig:overview}, \ourtool{} can individually or simultaneously integrate static analysis into the following three phases along the pipeline of model inference. 

\begin{itemize}[leftmargin=10pt]
\item \textbf{Prompting Phase.} Before model inference, static analysis can be integrated to provide relevant code contexts (\eg{} extracting code dependencies) for the input prompt.
\item \textbf{Decoding Phase.} During model inference, static analysis can be integrated to adjust the probability of tokens predicted by models on the fly.
\item \textbf{Post-processing Phase.} After model inference, static analysis can be integrated to filter and re-rank the model-generated code candidates. 
\end{itemize}
The integration in these three phases is theoretically orthogonal and can be  combined with each other by simultaneously applying them  during code completion.  In particular, \ourtool{} is customizable for selectively activating one or multiple specified integration strategies; and it is also extendable as new integration strategies of these three phases can be easily plugged into the framework. We then explain the integration strategies of each phase in detail. 

\subsection{Integration in Prompting Phase}\label{promptconstruction}
One key challenge of repository-level code completion is that the unfinished code might  depend on code located beyond the current file (\eg{} method invocations declared in other files of the repository). Therefore, given the unfinished code, static analysis can extract code dependencies from other files of the repository. The extracted cross-file code contexts can then be prepended to the current-file context in the prompt, providing hints to LLMs with potentially-relevant code ingredients. Specifically, our framework currently includes two granularities of code dependencies extracted by static analysis: file-level or token-level dependencies. 

\begin{itemize}[leftmargin=10pt, topsep=2pt]
\item \textbf{File-level Dependency Analysis.} This coarse-grained strategy focuses on file-level dependencies by analyzing the import statements of the current file. It is also a  common strategy used in existing repository-level code completion techniques (\eg{} RLPG~\cite{rlpg} and CoCoMIC~\cite{cocomic}). Specifically, given the unfinished code, we first leverage static analysis to extract the import statements in its current file and then extract all the classes and methods from these imported files. Given the limited input length of LLMs and to avoid overwhelming LLMs with massive contexts, we mainly organize the key information in a hierarchical way, \ie{} (i) for each module, we include its classes, (ii) for each class, we include its class signature, member variable names, and member methods, and (iii) for each method, we include its method signature. The ``\importcontext{}'' in Figure ~\ref{fig:overview} provides an example of file-level dependencies.

\item \textbf{Token-level Dependency Analysis.} This fine-grained strategy focuses on token-level dependencies that are valid at the completion position of unfinished code. In particular, given the unfinished code, we leverage static analysis to extract the list of method names and variable names that are valid at the completion position. To the best of our knowledge, directly incorporating such fine-grained dependencies into prompts has not been explored in previous repository-level code completion research. Different from the massive code contexts of file-level dependencies, token-level dependencies only include a precise token list into the prompt. The ``\apicontext{}'' in Figure ~\ref{fig:overview} provides an example of extracted  token-level dependencies. 
\end{itemize}

\subsection{Integration in Decoding Phase}
Static analysis can be integrated into the auto-regressive decoding procedure by on-the-fly adjusting the LLM-predicted token probabilities with static checking. 

Given the input token sequence in the current iteration $(x_1, x_2, \\ \ldots,x_n)$, the model generates a vector of $logits = (l_1, l_2, \ldots, l_K)$ for $K$ potential next tokens, where $K$ is the size of the vocabulary. In the auto-regressive greedy decoding procedure, the next token $x_{n+1}$ is decoded based on the top-1 probability computed with the \textit{softmax} and $argmax$ function:

\vspace{-4mm}
\begin{equation}
\label{eq2}
    x_{n+1} = \arg\max_{k}(\text{softmax}(logits))
\end{equation} 
\vspace{-3mm}

To incorporate static analysis into this procedure, we leverage static analysis to obtain all available APIs  (\eg{} the method names, variable names, and parameter names that have been defined) at the current position as the \textit{valid} tokens, and utilize the tokenizer to encode the first token and obtain a valid token id.
The vector $valid = (v_1, v_2, \ldots, v_K)$ indicates whether each token in the vocabulary is valid or not:
\begin{equation}
    v_k = 
    \begin{cases} 
    1 & \text{if the $k$-th token is valid} \\
    0 & \text{otherwise}
    \end{cases}
\end{equation}

Subsequently, the original logits vector $logits$ can be adjusted into $logits^+$, where
\vspace{-1mm}
\begin{equation}
    logits^+ = (l'_1, l'_2, \ldots, l'_K) = logits \oplus valid
\end{equation}
\vspace{-3mm}

$\oplus$ defines element-wise for vectors $logits$ and $valid$:

\vspace{-2mm}
\begin{equation}
    l'_k = 
    \begin{cases} 
    l_k & \text{if $v_k = 1$} \\
    -inf & \text{otherwise}
    \end{cases}
\end{equation}
\vspace{-2mm}

Equation ~\ref{eq2} can then be updated by replacing $logits$ with $logits^+$ :
\vspace{-3mm}
\begin{equation}
    x_{n+1} = \arg\max_{k}(\text{softmax}(logits^+))
\end{equation} 
\vspace{-3mm}

After being processed by \textit{softmax} function, the probability of all invalid tokens (whose prefixes previously are not returned by static analysis) is set as zero, thus being excluded in the current iteration of token generation.  
 
Figure ~\ref{fig:overview} illustrates how this integration strategy works. Compared to the prompting phase, integration in the decoding phase assigns more dominance to static analysis, as its returned information directly changes the generating probabilities of tokens. 

\subsection{Integration in Post-processing Phase}
In fact, incorporating static analysis in both prompting and decoding phases still cannot guarantee the final correctness of LLM-generated code, due to the nondeterminism and the limited interpretability of the model inference.
Therefore, leveraging static analysis in the post-processing phase to filter and re-rank the incorrect generated code can further improve the accuracy of LLM-based code completion. In our framework, we incorporate a straightforward integration strategy in the post-processing phase as follows. In particular, given the unfinished code, we first collect a ranked list of code candidates generated by LLMs with beam-search (\ie{} asking LLMs to generate top-k most possible code candidates); then for each generated code candidate, we put it back to the unfinished code and check whether the concatenated code can pass the static checking within the repository context, \ie{} no parsing or no compilation errors are reported by the static analyzer; lastly, the invalid code candidates are removed from the list and the first code candidate that can pass the static checking is returned as the final output. In addition, if all the code candidates fail the static checking, we return the original top-1 candidate generated by LLMs. Figure ~\ref{fig:overview} illustrates how the integration strategy works in the post-processing phase.

\section{Experiment Setup}
Based on our framework \ourtool{}, we investigate how static analysis can enhance LLM-based repository-level code completion by answering the following four research questions. 
\begin{itemize}[leftmargin=10pt, topsep=5pt]
\item \textbf{RQ1 (Individual Integration Strategy):} 
How does each static analysis integration strategy perform in LLM-based repository-level code completion?
\item \textbf{RQ2 (Combined Integration Strategies):} How do different combinations of integration strategies affect LLM-based repository-level code completion?
\item \textbf{RQ3 (Compared/Combined with RAG):} How do static analysis integration strategies perform when compared or combined with RAG in LLM-based repository-level code completion?
\item \textbf{RQ4 (Efficiency):} What are the online costs of different integration strategies in LLM-based repository-level code completion?
\end{itemize}

\subsection{Benchmark}
We choose the \cceval{}~\cite{cceval} benchmark for the following reasons. First, it is the latest repository-level code completion benchmark, constructed from GitHub projects spanning from March 5, 2023, to June 15, 2023. This timeframe helps mitigate data leakage issues by excluding training data from many existing code LLMs released before mid-2023. Second, \cceval{} is a more challenging benchmark  exclusively focusing on code completion with cross-file dependencies. As shown in previous work~\cite{cceval}, LLMs exhibit much poorer performance on \cceval{} than on other benchmarks such as RepoEval~\cite{repocoder}. Third, \cceval{} covers multiple programming languages, enabling experiments across diverse languages. 

\parabf{Studied Programming Languages.} We focus on repository-level code completion for two programming languages, \ie{} Java and Python, to cover both static and dynamic languages. Specifically, we adopt the Java and Python repository-level code completion tasks in the \cceval{} benchmark. Detailed statistics of the benchmark are provided in Table ~\ref{tab:dataset}. Note that the scale of the benchmark slightly reduces compared to the original \cceval{} (\eg{} from 239 Java repositories to 230 Java repositories) due to the removal of invalid data items from non-existent repositories or with non-existent modules imported.
In particular, the granularity of code completion tasks in \cceval{} is line level, \ie{} completing the current unfinished line within the repository-level context.

\begin{table}[h]
\centering
\caption{Benchmark statistics}
\label{tab:dataset}
\begin{adjustbox}{width=0.7\columnwidth}
\begin{tabular}{lrr}
\hline
Language           & Python  & Java \\ \hline
\# Repo. &  450 &  230         \\ 
\# Examples    &  2,460 & 2,075       \\
Avg. \# prompt lines    & 92   & 107       \\ 
Avg. \# prompt tokens    & 1,119   &  1,194      \\  
Avg. \# reference tokens   &  14  &  16      \\ \hline
\end{tabular}
\end{adjustbox}
\end{table}

\subsection{Studied Code LLMs}
We select three state-of-the-art code LLMs in our experiments, \ie{} \starcoder{}~\cite{starcoder}, \codellama{}~\cite{codellama}, and \deepseekcoder{}~\cite{deepseekcoder}, for the following  reasons. First, they are trained on data predating March 2023 and the repositories in \cceval{} are created between March 5 and June 15, 2023~\cite{cceval}, which can mitigate data leakage issues. Second, their input window size is at least 8,000 tokens, which can support the massive cross-file code contexts involved in repository-level code completion. Third, these code LLMs have been widely used in previous repository-level code completion work~\cite{cceval,repoformer,deepseekcoder} and shown advanced effectiveness. In particular, we mainly focus on the 6B-7B versions of these models given the cost-effectiveness balance. 

\subsection{Studied Baselines}
According to previous work~\cite{cceval}, we include the following two repository-level code completion baselines in our experiments:
\begin{itemize}[leftmargin=10pt, topsep=3pt]
\item \textbf{\infile{} Generation} is a default generation strategy including only the code preceding to the unfinished code into the prompt, which is the most common baseline widely included in repository-level code completion studies~\cite{cceval,repocoder,cocomic,RepoBench,repofusion}. 
\item \textbf{RAG} is another mainstream approach for enhancing LLM-based repository-level code completion~\cite{repocoder, repoformer}. We select \repocoder{}~\cite{repocoder} as a representative RAG-based technique. It retrieves the code similar to the current unfinished one from the repository and  concatenates it with the in-file context to form the prompt. 
\end{itemize}

\subsection{Metrics}
In line with previous work~\cite{cceval,repocoder}, we adopt the following four metrics to evaluate the effectiveness of different repository-level code completion techniques.  
\begin{itemize}[leftmargin=10pt, topsep=2pt]
    \item \textbf{Code Match.}  These metrics evaluate the similarity between the entire generated code and the ground truth: 
    \begin{itemize}[leftmargin=5pt, topsep=1pt]
    \item \textbf{Line Exact Match (Line EM)} measures the percentage of generated code that exactly matches the ground truth.
    \item \textbf{Line Edit Similarity (Line ES)} calculates the Levenshtein distance between the generated code and the ground truth~\cite{levenshtein1966binary}.
    \end{itemize}
    \item \textbf{Identifier Match.} These metrics assess the accuracy of generating correct APIs by comparing identifiers in the generated code with those in the ground truth.
    \begin{itemize}[leftmargin=5pt, topsep=1pt]
    \item \textbf{Identifier Exact Match (ID EM)} measures the percentage of generated code with the same identifiers as the ground truth. 
    \item \textbf{Identifier F1} is the F1 score of the identifiers in the generated code against the ground truth. 
    
    \end{itemize}
\end{itemize}

\subsection{Implementations}
\parabf{Prompt Construction.} 
The static analysis contexts are detailed in Section~\ref{promptconstruction}. For constructing retrieval contexts, we follow the settings in prior RAG-based code completion~\cite{repocoder}, which partitions the repository code using a sliding window with a line length of 20 and a sliding size of 10. We adopt lexical-based similarity, specifically Jaccard similarity~\cite{jaccard}, to calculate the similarity between different code snippets during retrieval.
We opt not to use deep learning-based retrievers (\eg{} CodeBERT~\cite{codebert}), as previous studies ~\cite{repocoder,cceval} show 
no significance effectiveness difference between lexical-based and deep learning-based retrievers in repository-level code completion. 
Furthermore, we include 3 retrieved code snippets into the prompt, as recent research~\cite{codesearch} indicates that 1 or 2 shots can already yield notable improvement, while 4 or more shots lead to limited increases or even decreases in performance. 
Given the limited input size of 8,000 tokens and the average prompt token length in the benchmark being less than 2,000 (\ie{} 1,194/1,119 for Java/Python), we truncate the in-file context to 2,000 tokens from left to right and truncate each cross-file context to 3,000 tokens. This ensures adequately and equally sized input windows for different prompting-phase integration techniques.
All contexts are subsequently combined into the prompt, as depicted in Figure~\ref{fig:overview}. 
Additionally, considering resource costs, our experiments currently set \repocoder{} to one iteration when integrated into our framework.

\parabf{Static Analysis Details.} In the prompting phase, we adopt the widely-used toolkit \textit{tree-sitter} ~\cite{treesitter} for Java and the internal \textit{ast} module for Python to parse the code into abstract syntax trees (AST) for import statement identification. In addition, we adopt well-established language servers (\ie{} Jedi ~\cite{jedi} for Python and \textit{Eclipse JDT} ~\cite{jdt-ls} for Java) to extract code elements and obtain the list of valid tokens. To reduce the online costs during code completion, we initially parse each repository to collect and index dependencies in advance. 
In the post-processing phase, we adopt the static analyzers (\eg{} \textit{Pylint}~\cite{pylint} for Python and native \textit{javac} for Java) to check the correctness of the generated code.

\parabf{Experimental Details.}
We acquire all models directly from their official open-source repositories.
We employ the default greedy-decoding strategy without sampling, since previous work~\cite{cceval} has illustrated the marginal differences between them in similar repository-level line completion scenarios.
In the post-processing phase, we use beam search to generate the top-3 predictions as candidates, balancing accuracy with limited computational resources. Additionally, we define the termination condition for the generation process as the output of the first newline token to measure the accurate time cost of completing the whole line.
We set the \texttt{max\_new\_tokens} parameter to 64 for all models.
All experiments are conducted on a server with the Ubuntu 20.04.6 LTS operating system and equipped with eight A800-80G GPUs.

\begin{table*}[]
\centering
\caption{Results of individual integration strategies. ``\textit{\si{}}''/``\textit{\sls{}}'' denotes the prompting-phase integration with file-level/token-level dependencies, ``\textit{\sd{}}'' denotes the decoding-phase integration, and ``\textit{\spo{}}'' denotes the post-processing-phase integration. The numbers in parentheses show the improvement over in-file generation.}
\label{tab:rq1}
\begin{adjustbox}{width=2\columnwidth}
\renewcommand{\arraystretch}{1.1}
\begin{tabular}{c|c|cc|cc|cc|cc}

\hline
\multirow{3}{*}{\textbf{LLM}} & \multirow{3}{*}{\textbf{Strategy}} & \multicolumn{4}{c|}{\textbf{Python}} & \multicolumn{4}{c}{\textbf{Java}} \\ \cline{3-10}
 &  & \multicolumn{2}{c|}{\textbf{Code Match}} & \multicolumn{2}{c|}{\textbf{Identifier Match}} & \multicolumn{2}{c|}{\textbf{Code Match}} & \multicolumn{2}{c}{\textbf{Identifier Match}} \\ \cline{3-10}
 &  & \textbf{Line EM} & \textbf{Line ES} & \textbf{ID EM} & \textbf{F1} & \textbf{Line EM} & \textbf{Line ES} & \textbf{ID EM} & \textbf{F1} \\ \hline
\multirow{5}{*}{\textbf{\begin{tabular}[c]{@{}c@{}}DeepSeek-Coder\\ 6.7B\end{tabular}}} & \textbf{\infile{}} & 9.02 & 62.51 & 15.45 & 48.00 & 26.7 & 73.38 & 34.36 & 58.62 \\
 & \textbf{\si{}} & \bestres{27.80(↑18.78)} & {\ul \textbf{72.64(↑10.13)}} & {\ul \textbf{36.75(↑21.30)}} & {\ul \textbf{64.94(↑16.94)}} & {\ul \textbf{44.24(↑17.54)}} & {\ul \textbf{80.33(↑6.95)}} & {\ul \textbf{53.25(↑18.89)}} & {\ul \textbf{73.19(↑14.57)}} \\
 & \textbf{\sls{}} & 24.84(↑15.82) & 69.89(↑7.38) & 33.94(↑18.49) & 61.70(↑13.70) & 38.65(↑11.95) & 76.95(↑3.57) & 47.18(↑12.82) & 68.42(↑9.80) \\
 & \textbf{\sd{}} & 16.02(↑7.00) & 63.92(↑1.41) & 23.37(↑7.92) & 54.30(↑6.30) & 33.11(↑6.41) & 74.34(↑0.96) & 41.45(↑7.09) & 64.49(↑5.87) \\
 & \textbf{\spo{}} & 11.54(↑2.52) & 63.39(↑0.88) & 17.36(↑1.91) & 49.44(↑1.44) & 30.36(↑3.66) & 73.63(↑0.25) & 38.07(↑3.71) & 60.96(↑2.34) \\ \hline
\multirow{5}{*}{\textbf{\begin{tabular}[c]{@{}c@{}}StarCoderBase\\ 7B\end{tabular}}} & \textbf{In-file} & 6.38 & 60.82 & 13.17 & 45.53 & 23.61 & 72.22 & 32.24 & 56.95 \\
 & \textbf{\si{}} & {\ul \textbf{23.58(↑17.20)}} & {\ul \textbf{69.78(↑8.96)}} & {\ul \textbf{32.20(↑19.03)}} & {\ul \textbf{61.35(↑15.82)}} & {\ul \textbf{40.19(↑16.58)}} & {\ul \textbf{78.90(↑6.68)}} & {\ul \textbf{50.60(↑18.36)}} & {\ul \textbf{71.23(↑14.28)}} \\
 & \textbf{\sls{}} & 20.28(↑13.90) & 66.31(↑5.49) & 28.62(↑15.45) & 56.49(↑10.96) & 34.51(↑10.90) & 75.72(↑3.50) & 44.19(↑11.95) & 66.09(↑9.14) \\
 & \textbf{\sd{}} & 13.09(↑6.71) & 62.21(↑1.39) & 20.73(↑7.56) & 52.13(↑6.60) & 30.89(↑7.28) & 73.75(↑1.53) & 39.90(↑7.66) & 63.37(↑6.42) \\
 & \textbf{\spo{}} & 7.73(↑1.35) & 60.32(↓0.50) & 13.70(↑0.53) & 45.78(↑0.25) & 26.12(↑2.51) & 72.46(↑0.24) & 34.89(↑2.65) & 58.20(↑1.25) \\ \hline
\multirow{5}{*}{\textbf{\begin{tabular}[c]{@{}c@{}}CodeLlama\\ 7B\end{tabular}}} & \textbf{In-file} & 7.20 & 60.69 & 13.94 & 45.62 & 25.59 & 72.85 & 33.45 & 57.97 \\
 & \textbf{\si{}} & {\ul \textbf{23.94(↑16.74)}} & {\ul \textbf{69.77(↑9.08)}} & {\ul \textbf{33.01(↑19.07)}} & {\ul \textbf{61.18(↑15.56)}} &{\ul \textbf{41.25(↑15.66)}} & {\ul \textbf{78.79(↑5.94)}} & {\ul \textbf{50.99(↑17.54)}} & {\ul \textbf{71.12(↑13.15)}} \\
 & \textbf{\sls{}} & 21.46(↑14.26) & 67.34(↑6.65) & 30.33(↑16.39) & 58.11(↑12.49) & 33.35(↑7.76) & 74.95(↑2.10) & 42.46(↑9.01) & 64.82(↑6.85) \\
 & \textbf{\spo{}} & 8.87(↑1.67) & 60.47(↓0.22) & 14.92(↑0.98) & 45.84(↑0.22) & 27.66(↑2.07) & 72.63(↓0.22) & 35.47(↑2.02) & 58.73(↑0.76) \\ \hline

\end{tabular}
\end{adjustbox}
\end{table*}
\section{Results}
\subsection{RQ1: Individual Integration Strategy}
Table~\ref{tab:rq1} presents the effectiveness of each individual integration strategy. In particular, we skip the decoding-phase integration on \codellama{},  as its  SentencePiece~\cite{sentencepiece} tokenizer escapes whitespace and the beginning of each word with a meta symbol ``\_''(U+2581). Therefore, each token is encoded with the meta symbol before it (\eg{} the first token of ``sendMessage'' will be ``\_send'' instead of ``send''); As only ``\_send'' is reserved during decoding, there will be a whitespace before each token after detokenization (\eg{} `` send Message''). The whitespace cannot be directly removed as it is hard to distinguish whether it is whitespace or the beginning of a word. Based on the table, we have the following observations.

\parabf{\textit{Comparison among different strategies.}} Overall, we find that all the static analysis integration strategies can individually improve the basic capabilities of three studied LLMs in repository-level code completion. For example, compared to in-file generation, the studied LLMs achieve 1.35 - 18.78 line EM improvements on Python and 2.07 - 17.54 line EM improvements on Java.

In particular, we find that integrating static analysis in the prompting phase (both file-level dependency analysis and token-level dependency analysis) can achieve much larger improvements than integration in other phases; meanwhile integrating in the post-processing phase achieves the smallest improvements. 
For example, for \deepseekcoder{} on Python, integrating static analysis in the prompting phase leads to 18.78/15.82 improvements (file-level/token-level dependency analysis) in Line EM, while integration in decoding or post-processing phases only has 7.00 or 2.52 improvements in Line EM.
One reason might be that including the relevant tokens before inference can better guide LLMs to be aware of these tokens and assign higher probabilities to them during the following inference. 
Figure~\ref{fig:prompt>post} shows an example that the prompting-phase integration can help \deepseekcoder{} generate the correct code with the API name but the decoding-phase integration or the post-processing integration cannot. 
As shown in the figure, when prompted with only the in-file context, the model tends to predict the local API ``normalize''. 
When utilizing the decoding-phase integration strategy, although it retains the ``strip'' token, the model does not assign enough probability (only 0.06) to it, which is still surpassed by the token ``normalize'' as it is also valid and with higher probability.

\begin{figure}[htb]
	\centering
  \vspace{1mm}	\includegraphics[width=1.05\columnwidth]{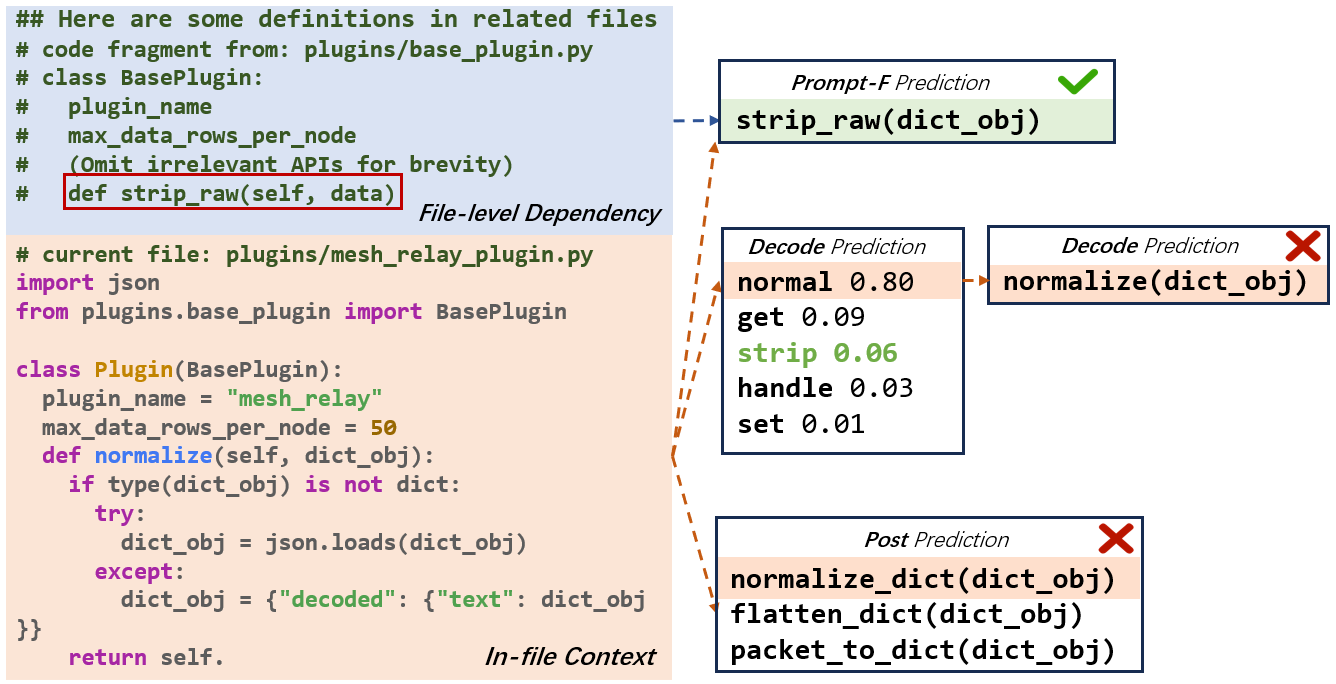}
	\caption{Comparison case of three strategies} 
	\label{fig:prompt_post}
\end{figure}

In addition,  prompting phase with file-level dependency analysis is the best individual integration strategy. The main reason why file-level dependencies outperform token-level dependencies might be that  (i) the file-level contains more comprehensive contexts regarding the usage of APIs,  and (ii) putting the cross-file context of the original code hierarchies at the beginning of the current file, aligns better with the data layout during LLM pre-training. Therefore, although file-level dependency analysis and token-level dependency analysis can both provide the correct API names in the prompt, it is easier for LLMs to utilize the relevant tokens displayed within file-level dependencies. 

\parabf{\textit{Generalization on different languages and LLMs.}} The observations above can be generalized on different languages and LLMs, as we observe consistent trends on all studied languages and LLMs. In particular, for both static and dynamic languages, integrating static analysis into any phase of code completion can improve the LLM-based repository-level code completion, while the prompting phase with file-level dependency analysis is always the best one with the highest improvements.

\finding{Integrating static analysis in any phase can improve repository-level code completion for all studied LLMs on both static and dynamic languages. Among them, integration in the prompting phase (especially with file-level dependencies) can achieve the substantially larger improvements than other phases, while integration in the post-processing phase achieves the smallest improvements.}
\vspace{-5mm}

\begin{table*}[]
\centering
\caption{Results of combined integration strategies.  Numbers in parentheses show  at least improvement over the individual one.}
\label{tab:rq2}
\begin{adjustbox}{width=2\columnwidth}
\renewcommand{\arraystretch}{1.1}
\begin{tabular}{c|c|cc|cc|cc|cc}

\hline
\multirow{3}{*}{\textbf{LLM}} & \multirow{3}{*}{\textbf{Strategies}} & \multicolumn{4}{c|}{\textbf{Python}} & \multicolumn{4}{c}{\textbf{Java}} \\ \cline{3-10}
 &  & \multicolumn{2}{c|}{\textbf{Code Match}} & \multicolumn{2}{c|}{\textbf{Identifier Match}} & \multicolumn{2}{c|}{\textbf{Code Match}} & \multicolumn{2}{c}{\textbf{Identifier Match}} \\ \cline{3-10}
 &  & \textbf{Line EM} & \textbf{Line ES} & \textbf{ID EM} & \textbf{F1} & \textbf{Line EM} & \textbf{Line ES} & \textbf{ID EM} & \textbf{F1} \\ \hline
\multirow{4}{*}{\textbf{\begin{tabular}[c]{@{}c@{}}DeepSeek-Coder\\ 6.7B\end{tabular}}} & \textbf{\si{} +  \sd{}} & 28.01(↑0.21) & 70.83(↓1.81) & 36.38(↓0.37) & 63.98(↓0.96) & {\ul \textbf{46.46(↑2.22)}} & {\ul \textbf{80.68(↑0.35)}} & {\ul \textbf{55.37(↑2.12)}} & {\ul \textbf{75.18(↑1.99)}} \\
 & \textbf{\si{} + \spo{}} & {\ul \textbf{28.74(↑0.94)}} & {\ul \textbf{72.40(↓0.24)}} & {\ul \textbf{36.91(↑0.16)}} & {\ul \textbf{65.04(↑0.10)}} & 45.93(↑1.69) & 80.49(↑0.16) & 54.80(↑1.55) & 73.79(↑0.60) \\
 & \textbf{\sls{} + \sd{}} & 24.51(↓0.33) & 68.08(↓1.81) & 32.80(↓1.14) & 60.55(↓1.15) & 39.04(↑0.39) & 76.53(↓0.42) & 47.47(↑0.29) & 68.44(↑0.02) \\
 & \textbf{\sls{} + \spo{}} & 26.95(↑2.11) & 70.05(↑0.16) & 35.45(↑1.51) & 61.99(↑0.29) & 40.34(↑1.69) & 76.97(↑0.02) & 48.48(↑1.30) & 69.01(↑0.59) \\ \hline
\multirow{4}{*}{\textbf{\begin{tabular}[c]{@{}c@{}}StarCoderBase\\ 7B\end{tabular}}} & \textbf{\si{} +  \sd{}} & 23.98(↑0.40) & 68.20(↓1.58) & 31.42(↓0.78) & 60.40(↓0.95) & {\ul \textbf{42.89(↑2.70)}} & {\ul \textbf{79.24(↑0.34)}} & {\ul \textbf{53.25(↑2.65)}} & {\ul \textbf{73.56(↑2.33)}} \\
 & \textbf{\si{} + \spo{}} & {\ul \textbf{25.09(↑1.51)}} & {\ul \textbf{70.10(↑0.32)}} & {\ul \textbf{32.94(↑0.74)}} & {\ul \textbf{61.92(↑0.57)}} & 41.73(↑1.54) & 78.89(↓0.01) & 51.76(↑1.16) & 71.20(↓0.03) \\
 & \textbf{\sls{} + \sd{}} & 20.81(↑0.53) & 64.62(↓1.69) & 28.53(↓0.09) & 55.57(↓0.92) & 35.04(↑0.53) & 75.46(↓0.26) & 44.82(↑0.63) & 66.38(↑0.29) \\
 & \textbf{\sls{} + \spo{}} & 22.34(↑2.06) & 66.53(↑0.22) & 29.58(↑0.96) & 56.54(↑0.05) & 37.11(↑2.60) & 76.42(↑0.70) & 46.41(↑2.22) & 67.49(↑1.40) \\ \hline
\multirow{2}{*}{\textbf{\begin{tabular}[c]{@{}c@{}}CodeLlama\\ 7B\end{tabular}}} & \textbf{\si{} + \spo{}} & {\ul \textbf{25.22(↑1.28)}} & {\ul \textbf{69.27(↓0.50)}} & {\ul \textbf{33.31(↑0.30)}} & {\ul \textbf{60.39(↓0.79)}} & {\ul \textbf{43.28(↑2.03)}} & {\ul \textbf{79.05(↑0.26)}} & {\ul \textbf{52.58(↑1.59)}} & {\ul \textbf{71.84(↑0.72)}} \\
 & \textbf{\sls{} + \spo{}} & 24.09(↑2.63) & 67.76(↑0.42) & 32.03(↑1.70) & 58.75(↑0.64) & 36.14(↑2.79) & 75.45(↑0.50) & 44.82(↑2.36) & 66.11(↑1.29) \\ \hline

 \end{tabular}
\end{adjustbox}
\end{table*}
\begin{figure*}[htp]
	\centering
    \vspace{2mm}	\includegraphics[width=2.1\columnwidth]{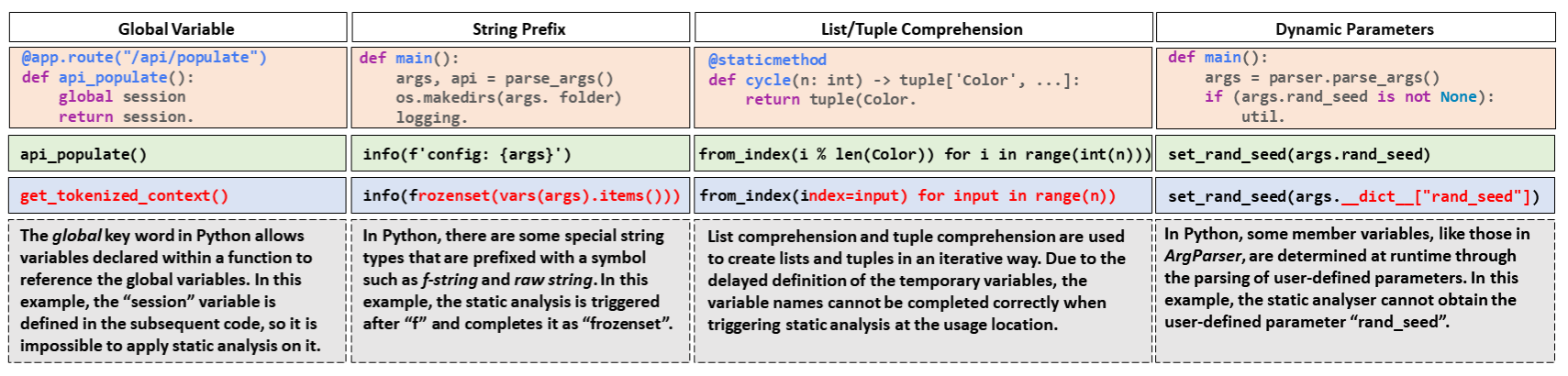}
	\caption{Bad case categories in decoding-phase integration (\colorbox{myorgange}{prompt}, \colorbox{mygreen}{ground truth}, \colorbox{myblue}{model prediction}, \colorbox{mygray}{explanation})} \label{fig:badcase}
 \vspace{-2mm}
\end{figure*}

\subsection{RQ2: Combined Integration Strategies}
Table~\ref{tab:rq2} presents the effectiveness when multiple integration strategies are combined. Although \ourtool{} theoretically supports the combination between any two strategies, our experiments skip the combination of decoding-phase integration and post-processing-phase integration, given the unacceptably expensive time and memory costs of their combination. For example, it takes \textbf{24} seconds to infer one data item on Java for \deepseekcoder{} when combining decoding-phase integration and post-processing-phase integration, as the latency can be amplified when the decoding-phase integration meets the bream search required by post-processing integration. Based on the tables, we have the following observations.

\parabf{\textit{Improvements from combining multiple strategies.}} Overall, combining any two integration strategies can further improve the accuracy of LLM-based code completion. For example, combining prompting-phase integration with either decoding-phase integration or post-processing-phase integration can both improve the original prompting-phase integration with 0.21 and 0.94 increase in Line EM. 
The only exception occurs when combining the decoding-phase integration with the prompting-phase integration on Python, where there is even a marginal decrease (\eg{} for \deepseekcoder{} on Python). The reason might be the limitations in analyzing dynamic language, which we would further discuss in the later paragraph ``Different results on different languages''. 
In addition, compared to the improvement of individual integration strategy over the in-file generation, the additional improvements from combination are relatively marginal, \eg{} the highest observed increment in \lineem{} is within 3.

\begin{figure*}[htb]
	\centering 
 \includegraphics[width=1.9\columnwidth]{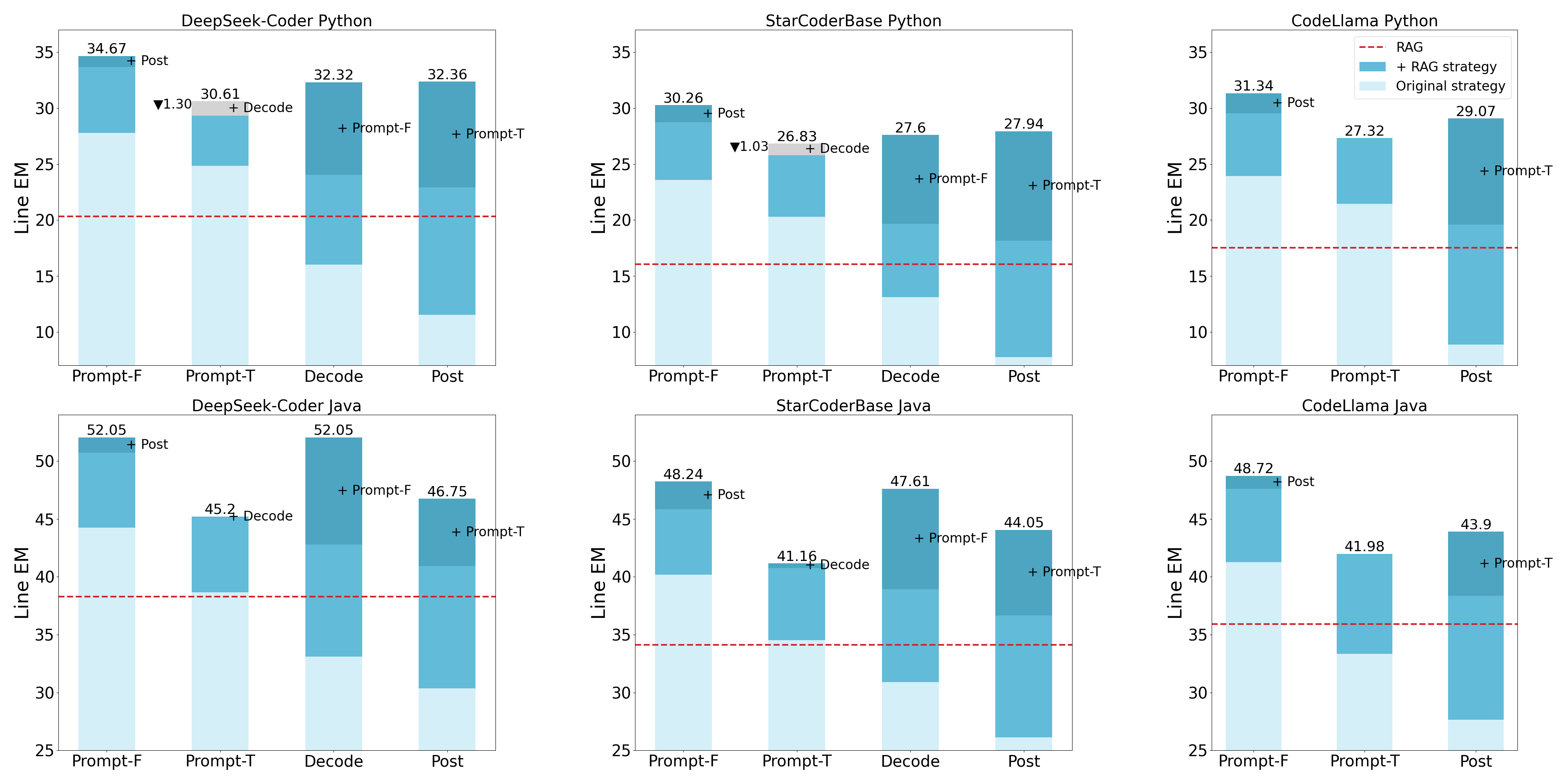}
	\caption{Combining \rg{} and static analysis integration strategies} 
	\label{fig:rq3}
\end{figure*}

\parabf{\textit{Complementarity between different strategies.}} The improvements of combining different integration strategies actually vary, indicating the different complementarity between integration strategies. In particular, we can observe the smallest complementarity between the prompting-phase integration with token-level dependencies and the decoding-phase integration, as their combination leads to at most 0.53 improvements in \lineem{}. The main reason might be that both token-level dependencies and the decoding-phase integration are built on the list of valid tokens (\eg{} the active method names or variable names), which share quite overlapped hints to the models. As a result, the benefits from combining them can be limited. In addition, we can observe a rather stable complementarity between post-processing integration and other phases.

\parabf{\textit{Different results on different languages.}} Interestingly, we can observe that the complementarity between integration strategies is different on static language (Java) and dynamic language (Python). Especially for the decoding-phase integration, it exhibits larger improvements when combined with other strategies on Java than on Python. In particular, for \deepseekcoder{} on Python, \lineem{} of the token-level dependencies even drops (\ie{} decreasing from 24.84 to 24.51) after combined with the decoding-phase integration.

\underline{\textit{Bad cases caused by imprecise static analysis.}} We further perform a qualitative analysis by manually inspecting the bad cases, and find that the observation above is mainly caused by the imprecise static analysis results originating from the inherent challenges of statically analyzing dynamic languages. In particular, we further summarize the bad cases that are caused by the imprecise analysis in Python into the four categories, each of which is illustrated with an example in Figure~\ref{fig:badcase}. As shown in Figure~\ref{fig:badcase}, the syntactic characteristics (\eg{} ``String Prefix'') and dynamic characteristics (\eg{} ``Dynamic Parameters'') in Python make it challenging for static analysis to return accurate results. As a result, static analysis can miss the correct tokens or introduce massive unrelated tokens during the decoding phase, which would mislead the model-predicted results by incorrectly adjusting the decoding logits.

\finding{Combining multiple integration strategies can bring further improvements for LLM-based repository-level code completion. The complementarity between  integration strategies is different: (i) the decoding-phase one and prompting-phase one with token-level dependencies share the smallest complementarity due to their similar information sources, and (ii) the improvements from combining with decoding-phase integration is larger on Java than on Python due to the difficulties in statically analyzing dynamic languages.}
\vspace{-5mm}
\subsection{RQ3: Compared/Combined with RAG}\label{rq3}
Retrieval-Augmented Generation (RAG) is another mainstream category for enhancing LLM-based repository-level code completion. In RQ3, we first compare static analysis integration strategies with RAG, and then combine them to explore further improvements. Figure~\ref{fig:rq3} shows the \lineem{} of combining static analysis and RAG. Based on the figure, we can have the following observations.

\begin{figure}[htb]
	\centering
  \includegraphics[width=1.0\columnwidth]{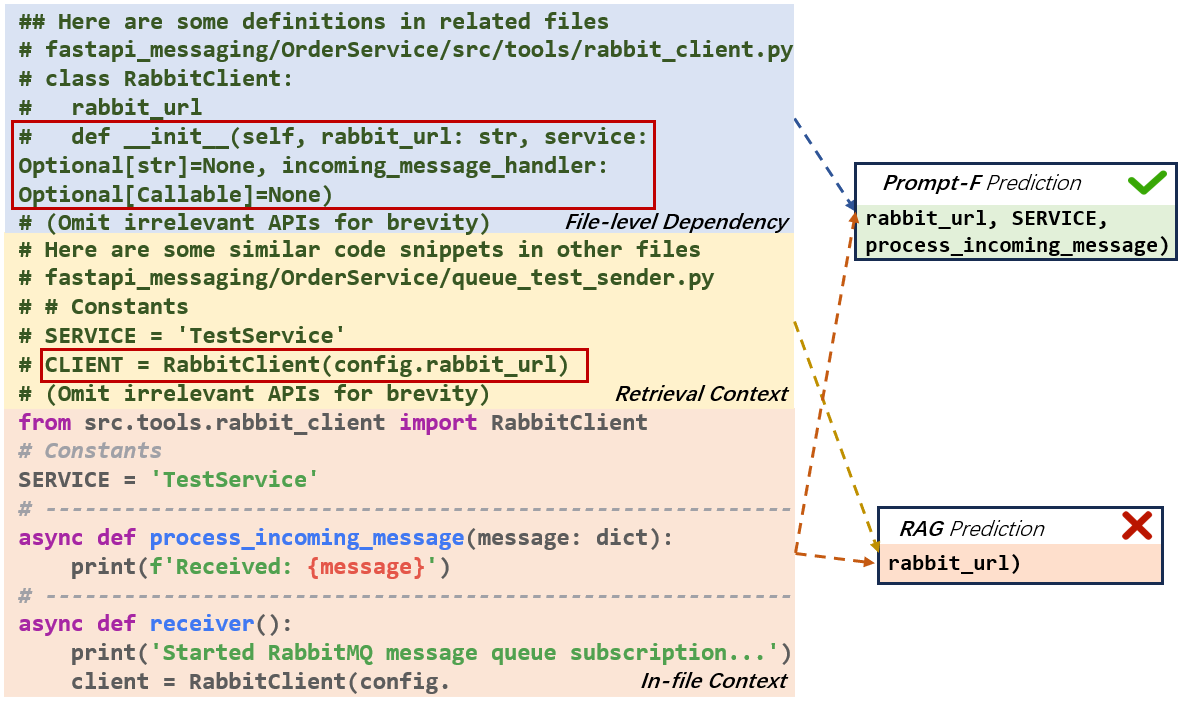}
	\caption{Comparison example of \rg{} and \si{}} 
	\label{fig:si_and_rg}
\end{figure}

\parabf{\textit{Compared with RAG.}} First, for the individual integration strategy, enhancing LLMs with static analysis in prompting phase (with either file-level dependencies or token-level dependencies) substantially outperforms enhancing LLMs with RAG. As shown in Figure~\ref{fig:rq3}, \lineem{} of both prompting-phase integration strategies are higher than RAG on most of the studied models for both Java and Python. However, RAG is more effective than integration strategies in decoding or post-processing phases. Second, for the combined integration strategies, the combination of any two static analysis integration strategies can substantially outperform RAG. To the best of our knowledge, it is the first time that static analysis integration has been shown to be superior to RAG in LLM-based repository-level code completion. 
Figure ~\ref{fig:si_and_rg} illustrates an example that static analysis integration outperforms RAG. In this example, the in-file context is slightly different from the RAG-retrieved context for a new function called \texttt{process\_incoming\_message}, which should be used as the third parameter when instantiating the class \texttt{RabbitClient}. However, the model mimics the API invocation in the similar code snippet retrieved by RAG, resulting in incorrect completion. In contrast, given ``File-level Dependency'' context, the model is provided with the full parameter information of the \texttt{\_\_init\_\_} function, and thus generates the correct completion.

\parabf{\textit{Combined with RAG.}} Overall, we can find that combining RAG with individual or multiple integration strategies can further substantially improve LLM-based repository-level code completion, to the best of our knowledge, even leading to the best accuracy on \cceval{} so far. For example, combining RAG with file-level dependencies and post-processing-phase integration achieves \textbf{34.67}/\textbf{52.05}  \lineem{} for \deepseekcoder{} on Python/Java, which is the highest accuracy on the \cceval{} benchmark (\ie{} the highest reported line EM with similar-scale models on \cceval{} is 21.35/21.69~\cite{cceval}). In addition, at most cases, combining RAG with file-level dependencies and post-processing  integration can achieve the best effectiveness. The results reveal the complementarity between static analysis and RAG, indicating that combining both techniques can enable more powerful LLM-based repository-level code completion.

\finding{Enhancing LLMs with static analysis achieves better effectiveness in repository-level code completion than enhancing LLMs with RAG. In addition, static analysis and RAG are complementary, and combining them can further make a better repository-level code completion technique, which can achieve the highest accuracy on the \cceval{} benchmark.}

\subsection{RQ4: Online Efficiency}~\label{sec:cost}
Table~\ref{tab:costs} shows the online time costs of 
different integration strategies.  The number calculates the average time costs per data item, including the online static analysis time and online model inference time. Based on the table, we can have the following observations.

\parabf{\textit{Efficiency comparison.}} Overall, among the individual integration strategies, integrating static analysis in the prompting phase is the most efficient way than other two phases (\ie{} decoding and post-processing phases). In particular,  integrating with file-level dependencies or token-level dependencies both takes less than 2 seconds, which introduces slightly more latency compared to the original in-file generation; however, there are noticeable increasing time costs in decoding-phase and post-processing-phase integration (\eg{} jumping to more than 5 - 10 seconds). The observations above can be caused by the following two reasons. First,  prompting-phase integration actually includes more contexts into the input, which thus introduces more overheads for inference compared to the basic in-file generation. In other words, longer inputs inherently take more time for inference, which can also explain why including file-level dependencies into prompt causes larger latency than token-level dependencies (as the former contains larger scale of context whereas the latter contains a list of available method or variable names). Second, the latency of decoding-phase and post-processing-phase integration is even larger, as (i) the decoding-phase integration requires iteratively invoking static analysis on the fly, and (ii) the post-processing integration requires beam-search to generate multiple candidates, causing non-trivial latency around 10s. 

Second, prompting-phase integration strategies exhibit comparable efficiency as the RAG-based approach, and the latter is quite stable on different languages and different code LLMs (\ie{} ranging from 1.5 to 2.0 seconds); on the other hand, decoding-phase and post-processing-phase integration are much more expensive than RAG. Therefore, combining RAG with prompting-phase integration strategies only induces acceptable increment of latency, but combing RAG with other integration strategies share the relatively-high costs as multiple integration strategies. 

\parabf{\textit{Trade-off between efficiency and accuracy.}} Overall, for the completion scenario of low latency requirements (such as human-in-the-loop coding assistant), combining prompting-phase static analysis and RAG is the best option for cost-effectiveness, which can achieve relatively high accuracy without introducing too much additional costs in inference. In addition, for the completion scenario of higher accuracy requirements, combining RAG with both prompting-phase static analysis and post-processing-phase static analysis can achieve the best accuracy. 

\finding{Integrating static analysis in the prompting phase is the most efficient way, which only introduces small additional time costs, while integrating static analysis in the decoding or post-processing phases introduces non-trivial additional latency, which is less efficient than RAG. In addition, combining prompting-phase static analysis and RAG is the best option for cost-effectiveness.}

\begin{table}[]

\centering
\caption{Online costs of different strategies (unit: s/item)}
\label{tab:costs}
\begin{adjustbox}{width=1.02\columnwidth}
\begin{tabular}{l|cc|cc|cc}
\hline
\multirow{2}{*}{\textbf{Strategies}} & \multicolumn{2}{c|}{\textbf{DeepSeek-Coder-7B}} & \multicolumn{2}{c|}{\textbf{StarCoderBase-7B}} & \multicolumn{2}{c}{\textbf{CodeLlama-7B}} \\ \cline{2-7}
 & \textbf{Python} & \textbf{Java} & \textbf{Python} & \textbf{Java} & \textbf{Python} & \textbf{Java} \\ \hline
\textbf{\infile{}} & 1.12 & 1.01 & 1.04 & 1.01 & 1.12 & 2.07 \\ \hline
\textbf{\rg{}} & 1.96(↑0.84) & 1.75(↑0.74) & 1.77(↑0.73) & 1.75(↑0.74) & 1.95(↑0.83) & 1.99(↓0.08) \\ \hline
\textbf{\si{}} & 1.70(↑0.58) & 1.67(↑0.66) & 1.69(↑0.65) & 1.67(↑0.66) & 1.85(↑0.73) & 1.90(↓0.17) \\
\textbf{\sls{}} & 1.53(↑0.41) & 2.53(↑1.52) & 1.40(↑0.36) & 2.53(↑1.52) & 1.53(↑0.41) & 1.39(↓0.68) \\
\textbf{\sd{}} & 5.72(↑4.60) & 9.61(↑8.60) & 4.33(↑3.29) & 9.61(↑8.60) & - & - \\
\textbf{\spo{}} & 9.78(↑8.66) & 4.23(↑3.22) & 6.24(↑5.20) & 4.23(↑3.22) & 7.48(↑6.36) & 4.80(↑2.73) \\ \hline
\textbf{\sid{}} & 5.51(↑3.81) & 9.70(↑8.03) & 5.87(↑4.18) & 9.70(↑8.03) & - & - \\
\textbf{\sip{}} & 10.50(↑8.80) & 11.33(↑9.66) & 9.03(↑7.34) & 11.33(↑9.66) & 9.00(↑7.15) & 6.67(↑4.77) \\
\textbf{\slsd{}} & 5.72(↑4.19) & 9.60(↑7.07) & 5.70(↑4.30) & 9.60(↑7.07) & - & - \\
\textbf{\slsp{}} & 9.60(↑8.07) & 4.72(↑2.19) & 7.86(↑6.46) & 4.72(↑2.19) & 9.40(↑7.87) & 5.22(↑3.83) \\ \hline
\textbf{\rsi{}} & 2.73(↑1.03) & 2.42(↑0.75) & 2.51(↑0.82) & 2.42(↑0.75) & 2.72(↑0.87) & 2.71(↑0.81) \\
\textbf{\rsls{}} & 2.36(↑0.83) & 1.93(↓0.60) & 2.16(↑0.76) & 1.93(↓0.60) & 2.33(↑0.80) & 2.17(↑0.78) \\
\textbf{\rsd{}} & 6.00(↑0.28) & 9.20(↓0.41) & 6.62(↑2.29) & 9.20(↓0.41) & - & - \\
\textbf{\rsp{}} & 11.48(↑1.70) & 6.43(↑2.20) & 9.71(↑3.47) & 6.43(↑2.20) & 7.90(↑0.42) & 7.06(↑2.26) \\
\textbf{\rsid{}} & 6.79(↑1.28) & 9.30(↓0.40) & 6.73(↑0.86) & 9.30(↓0.40) & - & - \\
\textbf{\rsip{}} & 12.75(↑2.25) & 8.38(↓2.95) & 11.35(↑2.32) & 8.38(↓2.95) & 11.54(↑2.54) & 9.25(↑2.58) \\
\textbf{\rslsd{}} & 6.34(↑0.62) & 9.70(↑0.10) & 9.54(↑3.84) & 9.70(↑0.10) & - & - \\
\textbf{\rslsp{}} & 12.17(↑2.57) & 6.94(↑2.22) & 9.23(↑1.37) & 6.94(↑2.22) & 10.52(↑1.12) & 7.63(↑2.41) \\ \hline

\end{tabular}
\end{adjustbox}
\end{table}

\section{Implications}
We then discuss the implications for future research. 

\parabf{More flexible integration strategies for static analysis to address its precision limitations.} Our findings reveal that the imprecise analysis results returned by static analysis (especially for dynamic language) sometimes can restrict the improvements from static analysis on LLM-based repository-level code completion. However, the current integration strategies in decoding phase or post-processing phase assign the dominate importance to the static analysis (\eg{} the tokens or statements that are identified as invalid would be directly filtered out). Therefore, to address this issue, it is important to design more flexible integration strategies that can tolerant the imprecise analysis results of static analysis (\eg{} deciding whether adopt the results returned by static analysis or not based on the confidence of LLMs), so as to mitigate the negative impacts from imprecise static analysis. 

\parabf{More advanced baselines for future evaluation.} Our findings demonstrate that combining multiple static analysis integration strategies with RAG-based approach achieves the best accuracy for repository-level code completion, which can serve as the new state-of-the-art baseline for LLM-based repository-level code completion. In addition, our findings also recommend different integration strategies for the application scenarios with different latency requirements, \ie{} prompting-phase integration strategies for the best cost-effectiveness whereas prompting-phase with post-processing-phase for the best accuracy but less efficient. 

\parabf{More efficient integration strategies.} Our findings show that some integration strategies for static analysis would introduce non-trivial costs during the model inference. Therefore, we call for the awareness of the efficiency issues when integrating static analysis into LLM-based code completion, especially for the latency introduced in the decoding and the post-processing phases. For example, one potential solution is to design  adaptive strategy for selectively invoking static analysis during the decoding phase instead of integrating static analysis in each iteration of decoding. 

\parabf{More diverse representation of dependencies in prompt.} Given the superiority of file-level dependencies over token-level dependencies in our findings, we believe that how to represent the dependencies extracted by static analysis in the prompt is essential for the accuracy of the code completion. In addition, the impact of dependency representation might also be associated with how the dependencies are organized in the model pre-training phase. Currently there is rather limited exploration on how to represent the dependencies of static analysis in the prompt, which can be very important future work to systematically explore more diverse dependency representation for repository-level code completion.

\section{Threats to Validity}
The validity of our findings might be threatened by the following issues. First, the findings might be impacted by the implementation limitations of static analysis. To mitigate this issue, we build our framework on the top of well-established and widely-used static analysis toolkit, \ie{} {tree-sitter}~\cite{treesitter}, Jedi~\cite{jedi}, and {Eclipse JDT}~\cite{jdt-ls}. Second, the generality of the findings can be specific to the benchmarks and code LLMs used in our experiments. To mitigate this issue, we include three state-of-the-art LLMs on both dynamic and static programming languages. Third, the code LLMs might have seen the data in the benchmark during pre-training, which can overestimate the effectiveness of code completion. To mitigate this data leakage issue, we select the latest code completion benchmark \cceval{} along with three code LLMs whose pretraining datasets are collected before the creation time of the benchmark.

\section{Related Work}
\subsection{Synergy between Static Analysis and LLM}
To mitigate the hallucination of LLM-based generation, there is an increasing body of research combining LLMs with static analysis techniques/tools so as to facilitate more powerful LLM-based solutions for software engineering tasks. For example, in automated program repair~\cite{DBLP:journals/corr/abs-2304-12743}, RepairAgent~\cite{Bouzenia2024RepairAgentAA} leverages static analysis to extract information for LLM-based repair; PyTy~\cite{Chow2024PyTyRS}  utilizes the type checker to verify whether each LLM-generated candidate accurately fixes static type errors; Repolit~\cite{Repilot} leverages completion engine to guide the patch generation on the fly. 
For bug detection, SkipAnalyzer~\cite{SkipAnalyzer} filters warnings from both LLM and static detection Infer; and GPTScan~\cite{GPTScan} employs static analysis to validate crucial variables and statements recognized by LLM, which allows to determine the existence of vulnerabilities with confidence. In addition, in software testing~\cite{wang2024software, DBLP:journals/corr/abs-2304-11686}, CodaMosa~\cite{codamosa} combines traditional search-based test generation with LLM-based test generation; TECO~\cite{TECO} trains transformer for test completion based on the semantic features extracted by static analysis; ChatTester~\cite{chattester} and TestPilot~\cite{DBLP:journals/corr/abs-2302-06527} leverage static analysis to prepare context information for the iterative LLM-based test code repair; AgentFL~\cite{AgentFL} builds a multi-agent system (some agents are enhanced by program analysis) to localize bugs within project-level contexts. 
The work above demonstrates the promise of combining static analysis and LLMs in different tasks.

\subsection{LLM-based Code Completion}
To date, many efforts have been dedicated to the LLM-based code completion, including code LLM training~\cite{deepseekcoder, starcoder, starcoder2, codegen, codellama, codex, Magicoder}, prompting engineering~\cite{SCoTs, CoTs, Zhou2022LeasttoMostPE}, benchmarks~\cite{RepoBench, cceval, HumanEval, classeval, CodeXGLUE, MBPP, multiple}, and empirical studies at diverse perspectives~\cite{izadi2024language, ReposVul, Ciniselli2021AnES, Ciniselli2021AnES, Dam2023EnrichingSC,xinghu24, DBLP:conf/issta/ShiW0DH0023}. The recent two surveys~\cite{zheng2023survey, Zan2022LargeLM} summarize the progress in LLMs for code completion. In particular, repository-level code completion is a more practical and challenging code completion scenario, while RAG-based and static analysis-integrated techniques are two mainstream categories of existing LLM-based repository-level code completion techniques. 
For RAG-based ones~\cite{repoformer, ReACC, repocoder}, they mainly design different similarity strategies (\eg{} dual-encoder model~\cite{ReACC}) with different retrieval strategies (\eg{} selectively or iteratively triggering RAG~\cite{repoformer, repocoder}), which enhances LLMs by prompting with similar code in the repository. Our work focuses on the static analysis-integrated category, which is orthometric to these RAG-based ones; in addition, our framework is compatible to further combining RAG-based techniques, which has already been shown with promising effectiveness in our experiments. 
For static analysis-integrated one~\cite{mgd, repofusion, cocomic}, they mainly focus on integrating static analysis into one phase (\eg{} prompting phase or decoding phase)  of LLM-based repository-level code completion. In particular, \cocomic{} and \rlpg{} propose to extract dependent code from the repository and then include the extract contexts during prompting; \mgd{}~\cite{mgd} adjusts the token probabilities based on the results returned by static analysis during decoding. 

\section{Conclusion}
This work performs the first study on the static analysis integration in LLM-based repository-level code completion by investigating both the effectiveness and efficiency of static analysis integration strategies across different phases of code completion. Our findings show that integrating file-level dependencies in prompting phase performs the best while the integration in post-processing phase performs the worse. Additionally, we find different improvements from static analysis between dynamic language and static language as well as the improvements from strategy combiantion; Additionally, we find the complementarity between RAG and static analysis integration as well as their cost-effectiveness after combination.

\balance
\bibliographystyle{ACM-Reference-Format}
\bibliography{ref}

\end{document}